\documentclass[preprint,12pt,authoryear]{elsarticle}

\usepackage{amssymb}
\usepackage{titlesec}
\usepackage{multirow}
\usepackage[dvipsnames]{xcolor}

\titleformat{\paragraph}
{\normalfont\normalsize\it}{\theparagraph}{1em}{}
\titlespacing*{\paragraph}
{0pt}{3.25ex plus 1ex minus .2ex}{1.5ex plus .2ex}

\journal{Computational Materials Science}

\begin{document}

\begin{frontmatter}



\title{Unsupervised learning for structure detection in plastically deformed crystals}


\author[inst1]{Armand Barbot}

\affiliation[inst1]{organization={Universite Paris-Saclay, ONERA, CNRS},%Department and Organization
            addressline={Laboratoire d'etude des microstructures}, 
            city={Chatillon},
            postcode={92322}, 
            country={France}}

\author[inst1]{Riccardo Gatti}

%\affiliation[inst2]{organization={Department Two},%Department and Organization
%            addressline={Address Two}, 
%            city={City Two},
%            postcode={22222}, 
%            state={State Two},
%            country={Country Two}}

\begin{abstract}

Detecting structures at the particle scale within plastically deformed crystalline materials allows a better understanding of the occurring phenomena. 
While previous approaches mostly relied on applying hand-chosen criteria on different local parameters, these approaches could only detect already known structures.

We introduce an unsupervised learning algorithm to automatically detect structures within a crystal under plastic deformation. This approach is based on a study developed for structural detection on colloidal materials. This algorithm has the advantage of being computationally fast and easy to implement. We show that by using local parameters based on bond-angle distributions, we are able to detect more structures and with a higher degree of precision than traditional hand-made criteria.

\end{abstract}


\end{frontmatter}


\section{Introduction}
\label{sec:introduction}

%%       \citet{<label>} ==> Jones et al. (2015)
%%       \citep{<label>} ==> (Jones et al., 2015)

Molecular dynamics simulation is a powerful method allowing to simulate at the particle scale different materials such as colloidal systems \citep{boattini_unsupervised_2019}, glassy materials \citep{barbot_rejuvenation_2020,barbot_characterization_2020} or metallic nanocrystals \citep{amodeo_mechanical_2017}. To help in interpreting the simulations results, being able to determine the local structure at the particle-scale is essential. To do so, several approaches were developed, mainly relying on local order parameters to describe the surrounding environment of each particle (number of neighbours, angles formed with the neighbours, ...) to detect underlying substructures in the simulated atomistic sample. Among these methods, we can cite the Bond Orientational Order (BOO) parameter \citep{steinhardt_bond-orientational_1983}, the Common Neighbours Analysis (CNA) \citep{honeycutt_molecular_1987}, or the Bond-Angle Distribution (BAD) \citep{ackland_applications_2006}.  Such methods were applied with success to study several phenomena such as crystal nucleation \citep{sanz_out--equilibrium_2008}, melting \citep{noori_study_2015} or plasticity \citep{ackland_applications_2006,stukowski_structure_2012,amodeo_atomistic_2014}. However, they are mostly relying on hand chosen criteria and thus only works for already known structures. 

Recently, different approaches using machine learning to detect substructures from local order parameters were developed. The first models, which relied on supervised learning \citep{geiger_neural_2013,dietz_machine-learning_2017,boattini_neural-network-based_2018}, suffered from the same problem as the hand-chosen techniques: they could only be trained to find an expected structure in the studied systems. However, a recent work from Boattini $et$ $al$ \citep{boattini_unsupervised_2019} successfully applied an unsupervised learning method using the BOO parameter to automatically detect different structures within a colloidal material without relying on a priori knowledge of the underlying structures. This method also had the advantage of being easy to implement and computationally fast.

While the BOO parameter is able to discriminate between the different structures in crystalline materials, it does not perform well when the simulated system is under deformation \citep{lechner_accurate_2008}. Being able to automatically detect substructures within crystal under plastic deformation, including some previously undetected and unexpected case, would open the way to a better understanding of crystal plasticity at the atomic scale. This requires to find a local order parameter suitable to discriminate substructures at the atomistic level for elastically and plastically deformed systems.

 In this study, we present a method inspired by the paper from Boattini $et$ $al$ \citep{boattini_unsupervised_2019} to automatically study and detect the different substructures within a crystal under plastic deformation appearing at the atomistic scale. This approach relies on the BAD parameters designed for single crystals to describe the environment around each atom, as highlighted in \citep{ackland_applications_2006}. This local parameter was already used in previous structure detection approaches relying on bond-angle distributions, showing the ability to discriminate structures associated with crystal plasticity \citep{ackland_applications_2006,amodeo_atomistic_2014}.

Our method consists on extracting the most pertinent BAD parameters using an autoencoder neural networks \citep{rumelhart_learning_1986,dietz_machine-learning_2017} before applying two clustering models: K-means \citep{lloyd_least_1982,macqueen_methods_1967} and DBSCAN \citep{ester_density-based_1996}. It has the advantage of being computationally fast and easy to implement. We finally test our approach on a FCC single crystal of Nickel under plastic deformation.

\section{Methods}
\label{sec:methods}

This section is divided in three main parts. We first describe the system on which we apply our structure detection method. We then depict the BAD parameters we use to capture the local environment of each particle. The last section focuses on the algorithms we use to detect the local atomistic substructures in the system from the BAD parameters.

\subsection{System}
\label{subsec:system}

In this study, we perform Molecular Dynamics (MD) simulations using LAMMPS \citep{thompson_lammps_2022} to plastically deform a Ni FCC defect free single crystal. These simulations are then used to generate a dataset to apply our method, based on unsupervised learning, developed to detect the different structures emerging in the system during plastic deformation.

\begin{figure}[h!]
\centering
\includegraphics[width=11cm]{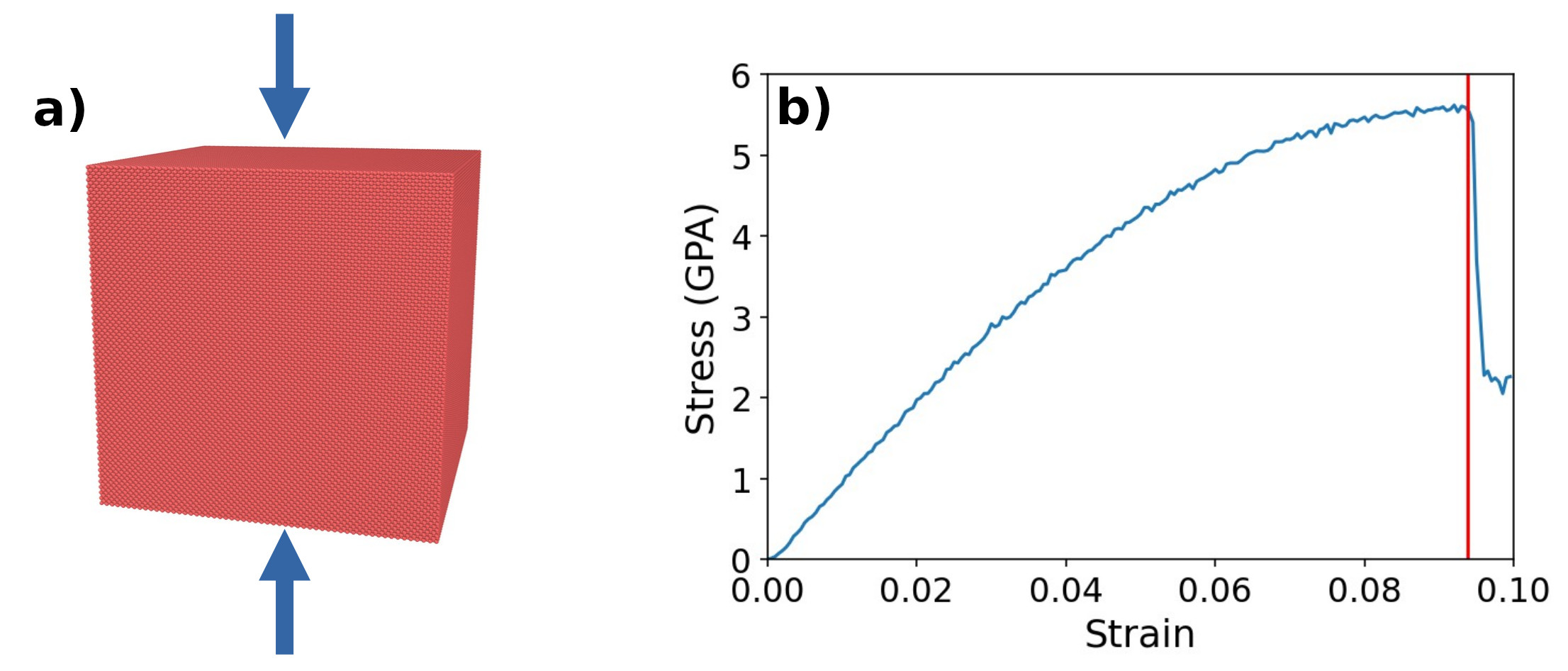}
\caption{(a) Snapshot made with Ovito \citep{stukowski_visualization_2009} of the system used in this study. It consists of a cubic nanoparticle of nickel with edge length equals to 20nm. The two arrows represent the uniaxial compression performed with a flat indenter along the $[100]$ direction. The stress-strain curve obtained from this compression is shown on (b). The vertical red line corresponds to occurence of the first plastic event.}
\label{fig:system}
\end{figure}

We first create a cubic sample with edge lengths equal to 20nm and free surfaces oriented along the $\langle 100 \rangle$ directions as shown on Fig \ref{fig:system} (a). The interatomic interaction linking the atoms is the EAM potential for Ni \citep{adams_self-diffusion_1989}. The system is then equilibrated for 5 ps at 5K using Nose-Hoover thermostat \citep{nose_unified_1984,hoover_canonical_1985}. A deformation is imposed using uniaxial compression with a flat indenter along the $[100]$ direction, with a strain rate of $10^8 s^{-1}$ at a temperature of 5K. The flat indenter is modeled as an infinite plane exerting a repulsive force on atoms defined by 
\begin{equation}
F(r)=-K(r-R)^2,
\end{equation}
with r being the atom position, R the plane position and K the force constant. Here we choose K=1000 eV/\AA $^3$. 
The temperature is maintained at 5K during the compression through the Nose-Hoover thermostat.

To determine when the first plastic event occurs during the deformation, we first measure on the fly the compressive stress applied on the system by the indenter using a method similar to the one used in \citep{amodeo_atomistic_2014} and looking for the first stress drop corresponding to the first plastic event. The stress-strain curve obtained from the compression is shown on Fig \ref{fig:system} (b), with the first plastic event represented by a vertical red line.

In our simulations, we observe the nucleation of Shockley partial dislocations with Burgers vectors: $\{ \frac{1}{6}[-\!1 \;  1 \; 2],\frac{1}{6}[1  \; -\!1 \;  2],\frac{1}{6}[-\!1  \; -\!1  \; 2],\frac{1}{6}[-\!1  \; -\!1  \; -\!2] \}$. The dislocation lines are thus accompanied by a stacking fault in the HCP structure \citep{hull_chapter_2011}.

It is important to underline that in this study, the system was deformed up to an applied strain of $0.1$, i.e., just after the yield point as we can see on Fig \ref{fig:system} (b). The substructures associated with plastic deformation in our dataset would then mostly correspond to dislocation lines, their interaction and the associated stacking fault. The results obtained by applying our model to a  system with larger deformation will be discussed in section \ref{sec:Results}.

\subsection{Bond Angle Distribution Parameter}
\label{subsec:BAD}

To describe the local structural environment around each particle, we measure the Bond-Angle Distribution (BAD) parameters for each of them, based on the method described in \citep{ackland_applications_2006}. 
For each atom i, we first extract its nearest neighbors $N_b(i)$ using the adaptive cutoff method as described in \citep{stukowski_structure_2012}. This method has the advantage of being parameter-free, so can be directly applied to any structure, while being just slightly slower than using a fixed cutoff \citep{stukowski_structure_2012}.

We define the bond-angle $\theta_{jik}$ with $j,k\in \{1..N_b(i)\}$ being two nearest neighbors of i. We then estimate $\{cos(\theta_{jik})\}$ for all the $N_b(i)(N_b(i)-1)/2$ neighbor pairs of atom i. 
From it, we describe nine ranges of $\{cos(\theta_{jik})\}$ over which the BAD parameters $\{\chi_l \}_{l \in \{0..8\}}$ are estimated as shown in Table \ref{Table:BAD}. These ranges are based on those from \citep{ackland_applications_2006} which were optimised to differentiate crystal structures. The idea here is to count the number of bond-angle cosines in each ranges to obtain the value of the corresponding $\{\chi_l \}_{l \in \{0..8\}}$.

\begin{table}[h!]
\caption{Definition of the nine bond angle cosines ranges over which the BAD parameters $\chi$ are estimated. $cos(\theta_{jik})$ is the cosines of the angle between $\bold{r}_{ij}$ and $\bold{r}_{ik}$, with $j,k$ being two nearest neighbors of the atom i. The ranges were taken from \citep{ackland_applications_2006} and were optimised to differentiate crystal structures.}
\vspace{0.2cm}
\centering
 \begin{tabular}{c c c c} 
 \hline
 BAD Parameter & Minimum $cos(\theta_{jik})$ & Maximum $cos(\theta_{jik})$ \\ [0.5ex] 
 \hline
 $\chi_0$ & -1.0 & -0.945 \\ 
 $\chi_1$ & -0.945 & -0.915 \\
 $\chi_2$ & -0.915 & -0.755  \\
 $\chi_3$ & -0.755 & -0.705\\
 $\chi_4$ & -0.705 & -0.195 \\
 $\chi_5$ & -0.195 & 0.195  \\
 $\chi_6$ & 0.195 & 0.245\\
 $\chi_7$ & 0.245 & 0.795 \\
 $\chi_8$ & 0.795 & 1.0  \\ [1ex] 
 \hline
 \end{tabular}
 \label{Table:BAD}
\end{table}

For instance, let us consider a case where a atom i has six bond-angle cosines. If two of them are in the range $[-1.0,-0.945]$ and four of them are in the range $[-0.945,-0.915]$, then for this atom, we will have $\chi_0=2$, $\chi_1=4$ and $\{\chi_l \}_{l \in \{2..8\}}=0$. 

Note that in this paper, the BAD parameters were calculated using the package Pyscal \citep{menon_pyscal_2019}.

\subsection{Unsupervised learning}
\label{subsec:UnsupervisedLearning}

From the knowledge of the BAD parameters for each atom, the aim of our method is to determine automatically the different structures present in the system. This part is done through unsupervised learning and can be decomposed on two main tasks. First we use an autoencoder to determine the good number of parameters to describe the local structures in our system. Then, we apply a perturbative method to the autoencoder to determine which parameters are the most effective. From this, we focus on the most effective parameters applying clustering and classification methods to extract the different substructures of our system.

Note that as the surface atoms can be easily filtered from their number of nearest neighbours as in \citep{amodeo_atomistic_2014}. %DOI:10.1080/21663831.2013.878884
, i.e. atoms for which $N_b<11$, we filter out the surface atoms using this criteria and focus on detecting the substructures on the  remaining atoms.

Furtermore, when training an unsupervised learning model for structural detection, it is essential to ensure that the dataset used for the model training contains at least the different sub-structures that should be detected. In fact, an insufficient training dataset would lead to miss of some structures when applying the trained model. In this paper, we made sure to generate a training dataset large enough to detect all the different substructures that are present in our MD simulations.

\subsubsection{Autoencoder}
\label{subsubsec:Autoencoder}

As explained above, the aim of this section is to extract the number of pertinent parameters to determine the different structures of our system. Indeed, out of the nine BAD parameters, some could be irrelevant or redundant to describe the local structural environment around each atom.
This task is here performed using an autoencoder based neural network to perform dimensionality reduction \citep{goodfellow_deep_2016,chen_collective_2018,boattini_unsupervised_2019}.

\begin{figure}[h!]
\centering
\includegraphics[width=6cm]{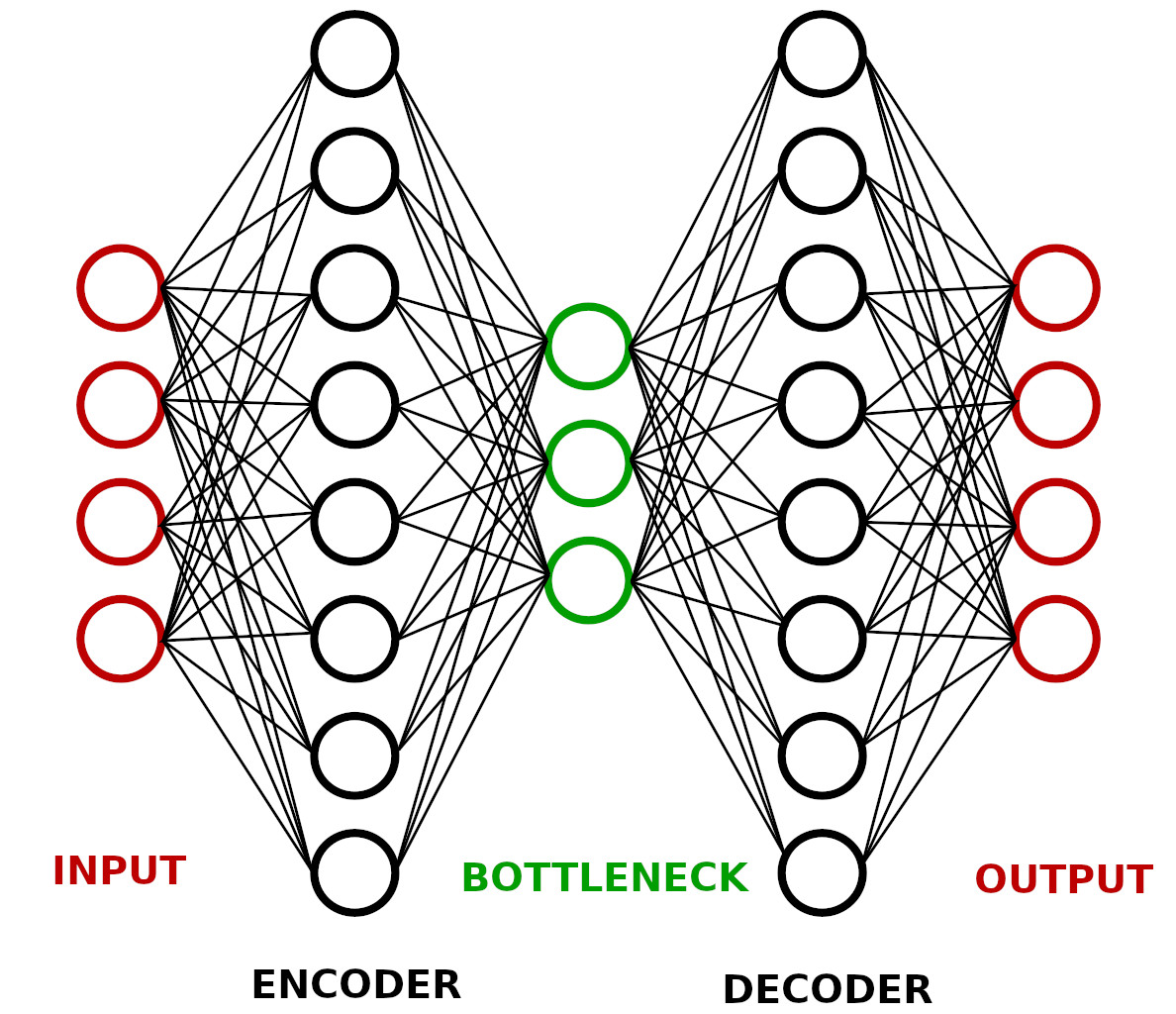}
\caption{Schematic representation of autoencoder neural network architecture for dimensional reduction. The neural network is trained in order to reproduce the input at the output while passing through a bottleneck of lower dimensions. The encoder network aims at finding a low-dimension representation of the input. The decoder network is trained to reconstruct the original input from the lower-dimension bottleneck.}
\label{fig:autoencoderschematics}
\end{figure}

Various methods exist to perform dimensionality reduction.  One of the most widely used methods is principal component analysis (PCA) \citep{hotelling_analysis_1933}, which consists of linearly combining the input parameters to create a new, smaller set of parameters in which most of the variation in the data is accounted for.  Although the PCA method may be sufficient in most dimensionality reduction tasks, in this study where we have 9 different parameters, we chose to focus  on autoencoders as they allow, copared to PCA, to more easily and reliably extract which input parameters are the most relevant for performing dimensionality reduction, and thus to obtain more information about the dataset. Furthermore, the autoencoders allow much complex combinations than the linear ones performed by the PCA. While an autoencoder can be slightly more complex to implement than PCA, the advantages outrun this added complexity. The implementation of autoencoders and the codes used for this study are available on open-source platforms \citep{barbot_armand_2023_7582668}.

in general, an autoencoder network can be divided in two parts: the encoder part and the decoder part as shown on Fig \ref{fig:autoencoderschematics}. The encoder aims at encoding the input into a lower dimension: the bottleneck. Then the decoder will try to reconstruct the input from the bottleneck. The output of the network will then be the reconstruction of the input done by the decoder from the bottleneck.

In practice, the input of the encoder is $\bold{\chi}(i)= \{\chi_l \}_{l \in \{0..8\}}(i) \in \mathbb{N}^9$, with $i \in [1..N]$, N being the number of atoms over which the training in performed. We train the encoder to reduce the difference between the input and the output for a bottleneck dimensions going from 1 to 9.

To determine the pertinent dimension of input, we measure for each bottleneck $N_{bneck}$ dimension the mean square error between the input $\bold{\chi}$ and the output $\bold{\hat{\chi}}$ as

\begin{equation}
E_{N_{bneck}}=\frac{1}{N} \sum^N_{i=1} \Vert \bold{\chi}(i)-\bold{\hat{\chi}}(i)\Vert.
\end{equation}
We then look from which dimension we have the minimum difference between the input and the output.

For more details about the implementation of the autoencoder, please refer to the paper \citep{boattini_unsupervised_2019} which served as a reference for this section.

\subsubsection{Relevant parameters identification}
\label{subsubsec:RelevantParameters}

After determining the relevant number of parameters to study the substructures in our system, we want to go beyond the black box of the neural network and determine which parameters are most pertinent for our study. To do so, we use the input perturbation method \citep{yao_forecasting_1998,scardi_developing_1999,gevrey_review_2003,olden_accurate_2004}.

The principle of this method is to take the trained autoencoder neural network with the optimal bottleneck dimension, and to perturbate one of the input parameters by increasing its value with a fixed amount. In this work, we increase the chosen input value by 10\% which was shown to give satisfactory results \citep{boattini_unsupervised_2019}.
 
 After perturbing one of the inputs, we look at how much this perturbation influenced the output by calculating:
 
 \begin{equation}
E^{pert}_k=\frac{1}{N} \sum^N_{i=1} \Vert \bold{\chi}(i)-\bold{\hat{\chi}}^{pert}_k(i)\Vert,
\label{Eq:PerturbatedError}
\end{equation}
with $k \in [0..8]$ being the index of the perturbed input, $\bold{\hat{\chi}}^{pert}_k$ the output after the perturbation and $\bold{\chi}$ the input without perturbation.

If the perturbed input $k$ was considered during the training as carrying no pertinent information to reconstruct the input at the output, the perturbation will not propagate to the output. This will lead to a very small value of $E^{pert}_k$. On the contrary, if the perturbed input k is necessary to obtain an output similar to the input, its perturbation will lead to a higher value of $E^{pert}_k$.

From this, we determine the relative importance $RI_k$ of the input k as:
 \begin{equation}
RI_k=\frac{\Delta E_k}{\sum_{j=0}^8 \Delta E_j},
\label{Eq:RelativeImportance}
\end{equation}
with $\Delta E_k=E^{pert}_k-E$ being the difference between the autoencoder mean square error after perturbation of the input k and without perturbation. From this relative importance index, we can determine the importance of each input during the autoencoder training.

\subsubsection{Clustering and classification} 
\label{subsubsec:ClusteringClassification}

In the last section, we described how to isolate the most relevant BAD parameters to study the local structure. In our system most of the atoms will be in two structures, FCC and HCP (in the stacking fault), which corresponds to close values of BAD parameters \citep{ackland_applications_2006}. While most structure detection approaches focus on the already known main crystalline structures, i.e. FCC, BCC and HCP \citep{stukowski_structure_2012}, the goal of our method is also to be able to automatically detect substructures without having an a priori knowledge about them.

To this end, we chose to apply two clustering methods on the relevant BAD parameters space, one to detect the main structures containing most of the atoms, and the second one to focus on the substructures containing few atoms but located far from each other in the parameter space. The clusters extracted by these two clustering approaches will then be combined to identify the different substructures of the studied system.

In the next sections, first we introduce the K-means clustering method \citep{macqueen_methods_1967,lloyd_least_1982} we use to detect the main structures. Then, we detail the Density-Based Spatial Clustering of Applications with Noise (DBSCAN) \citep{ester_density-based_1996} which allows to detect efficiently more isolated clusters. 

While K-means is able to efficiently determine the cluster of a new data point without having to re-apply the clustering over the whole data set, it is not the case for DBSCAN. We thus need to apply a classification algorithm over the clusters detected by DBSCAN. We thus present in the last part the logistic regression classifier we use to perform this task.

\paragraph{Kmeans} 
\label{paragraph:Kmeans}
 
\hspace{0.55cm}In order to detect the main substructures present in the system, we use the K-means clustering algorithm \citep{macqueen_methods_1967,lloyd_least_1982} as implemented in sci-kit learn \citep{pedregosa_scikit-learn_2011}. The principle of the method is the following: let us consider a set of N data points in the parameter space that we want to separate into K clusters. We name $\{x_i\}_{i\in[1..N]}$ the position of the data point in the parameter space and $\{\mu_j\}_{j\in[1..K]}$ the center of each cluster, commonly called "centroid". 

The aim of the K-means algorithm is to choose optimal centroids in order to minimize the inertia defined by 
 \begin{equation}
IN(K)=\sum^N_{i=1} min_{j\in[1..K]}(\Vert x_i-\mu_j \Vert^2).
\label{Eq:Inertia}
\end{equation}
If the points are located in clusters, the inertia will be minimized by placing the centroids at the center of each cluster.

While the K-means requires the number of clusters as an input, the optimal number of cluster can be obtained with the following procedure: we first apply K-means to our data set for varying number of clusters K and calculate the corresponding inertia $IN(K)$. In this paper, we did it for $K \in [1..10]$. Then we apply the elbow method to the function $IN(K)$ following the protocol described in \citep{salvador_determining_2004}. From this method, we can obtain the value of K from which the inertia stop decreasing with K, thus corresponding to the optimal number of clusters for K-means clustering.

After training the K-means algorithm over the data set for the optimal cluster number K, we obtain the centroids locations $\{\mu_j\}_{j\in[1..K]}$ of the different clusters. From this information, we can determine in which cluster is a new data point by looking at its closest centroid. This allows to classify new data depending on their cluster without having to retrain the algorithm each time.

The K-means method being only looking at the distance between the data points and the centroid, it does not perform well for elongated clusters or for clusters with irregular shapes. However, it performs very well if most of the data points are located in the same position in space. As the main crystalline structures are each associated with a specific position in the BAD parameter space, the K-means algorithm is well suited to detect the clusters associated with these structures.

\paragraph{DB Scan} 
\label{paragraph:DBScan}

\hspace{0.55cm}To be able to detect clusters with smaller number of points and non-isotripic shape, we apply another clustering method: the Density-Based Spatial Clustering of Applications with Noise (DBSCAN) \citep{ester_density-based_1996} as implemented in sci-kit learn \citep{pedregosa_scikit-learn_2011}. 

The principle of the algorithm is the following: we input two parameters: (1) the maximum distance between two points $l_{max}$ and (2) the minimum number of points per cluster $N_{min}$. 
In this method, two data points located at a distance $l<l_{max}$ are considered as neighbors. 
From it, the algorithm will class the data points into three categories: 

(i) if a data point has at least $N_{min}-1$ neighbors, it will then be considered as a "core point". 

(ii) if a data point has less than $N_{min}-1$ neighbors but at least one of them is a core point, it will be considered as a "border point". 

(iii) the remaining data points will be considered as "noise".

The core points and the border points are then regrouped together to form clusters, located at a distance $l_{cluster}>l_{max}$ from each other.
 In our system, many atoms can be associated with the same values of BAD parameters. As the DBSCAN only look at the distance between points, having many points located at the same spatial position will only slow down the clustering without improving the result. We thus apply the DBSCAN over the list of the unique spatial positions of our data set.
Finally, the BAD parameters being integers, we use for the distance parameter the value $l_{max}=1$. We also determined that $N_{min}=5$ gives satisfactory results for our data set.  

As this method is not influenced by the number of point at a given spatial position but only by the distance between the unique spatial positions in the data set, it would be able detect clusters associated with structure containing fewer atoms but isolated from the clusters corresponding to the main crystalline structures.  

Unlike K-means, it is not possible to directly add new points to determine from a pre-trained model in which cluster they are. To use DBSCAN with new points, it would require to add these new points to the training dataset and apply again the clustering method. This method, of course, requires that the training dataset contains at least the cluster at which the new point should be associated. 
This approach is however very time consuming, making the structure detection very slow. This is why we chose here to train a classifier in order to be able to quickly associate new points with a cluster.

\paragraph{Logistic regression classifier} 
\label{paragraph:LogisticRegression}

%In the opposite of K-means, if we want to know from DBSCAN in which cluster is a new data point, we would need to re-perform the clustering over the whole data set, plus the new data point. But this solution would be very inefficient and time consuming. One solution to solve this problem is to train a classification algorithm over the clusters detected by DBSCAN. 

\hspace{0.55cm}As detailed above, after detecting the different clusters in the training dataset with DBSCAN, we want to be able to efficiently determine at which cluster  a new added point belongs. To do so, we use a classification algorithm. 
The classification algorithm takes as an input the training data set and as the output the clusters found by DBSCAN. It will then decompose the parameter space into region; each associated with a cluster. When a new data point is added, it will be attributed a cluster depending on its location in the parameter space.
%\AB{It is here again important that the training dataset contains the cluster at which the new point should be associated. Otherwise, the new points will just be attributed to the closest cluster which would lead to an incorrect substructure detection.}

 In this study, we use the logistic regression classifier \citep{brzezinski_logistic_1999} implemented in scikit-learn as it simple, fast and gives satisfactory result when applied to the clusters which detected by DBSCAN in this system. More details about this classifier can be found in sci-kit learn \citep{pedregosa_scikit-learn_2011}.

\paragraph{Cluster combination} 
\label{paragraph:ClusterCombination}

\hspace{0.55cm}In the previous parts, we described two clustering methods. One to detect the main structures in the system: the K-means algorithm, the other one to detect the substructures containing few points, more isolated in the parameter space: the DBSCAN. Then, both K-means and the association of DBSCAN with the logistic regression classifier will separate the parameter space into regions corresponding to detected clusters. Thus, each data point will be associated with two clusters: the one detected by K-means, the other detected by DBSCAN.

In order to have a precise detection of the structures present in the system, we combine the results obtained from the two clustering methods.
 Finally, we obtain $N_{Kmeans} \cdot N_{DBSCAN}$ clusters, with $N_{Kmeans}$ and $N_{DBSCAN}$ corresponding to the number of clusters found by K-means and DBSCAN, respectively.

\section{Results and discussions}
\label{sec:Results}

In this section, we present the results obtained by following the procedure described in the previous sections. First we analyse the data obtained we get from the autoencoder: the relevant number of the BAD order parameter necessary to detect the substructures in our system as well as which of the $\bold{\chi}(i)= \{\chi_l \}_{l \in \{0..8\}}(i) \in \mathbb{N}^9$ parameters are the most significant. We then apply the clustering methods on the extracted parameter, later analysing their performances. Finally, we examine the different substructures detected by this method, comparing the outcomes with other structure detection methods.

\subsection{Autoencoder output analysis} 
\label{subsec:AutoencoderAnalysis}

After compressing our Ni crystal until the first plastic event occurs, as described in Section \ref{subsec:system}, we record at a regular time step interval the atomic positions of the simulated nano-object. We then measure BAD parameters for each atoms as detailed in Section \ref{subsec:BAD}, filtering out the surface atoms as explained in Section \ref{subsec:UnsupervisedLearning}. 
To train the autoencoder model, we used the BAD parameters extracted from a time step corresponding to an applied strain of $\Delta \epsilon=0.05\%$ after the first plastic event. This applied strain value was chosen to have snapshot of the system in which the nucleated dislocations propagate and begin to interact with each other, avoiding  dislocation lines annihilation at the surface.  To verify that no substructures were missed by our model, we applied our unsupervised learning approach to data constituted of other time steps than those chosen to train the model in order to check and to confirm that the training dataset contains at least the expected substructures that can be found when applying the model to the rest of the dataset.

%It turned out that computing the BAD parameters of a single time step is sufficient to train the autoencoder and to observe the appearance of different clusters. Here, the time step used to train our algorithm corresponds to an applied strain of $\Delta \epsilon=0.05\%$ after the first plastic event.

\begin{figure}[h!]
\centering
\includegraphics[width=8cm]{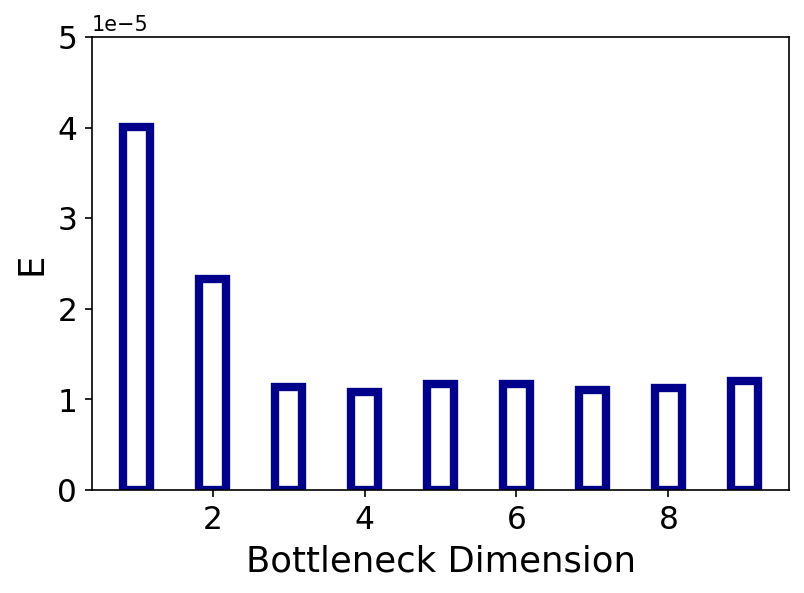}
\caption{Means square error $E$ between the input and the output for different bottleneck dimensions. The error reaches a plateau for a bottleneck dimension of $N=3$ which corresponds to the relevant number of BAD parameters sufficient to reconstruct the input data.}
\label{fig:bottleneck}
\end{figure}

Fig \ref{fig:bottleneck} shows the mean square error between the autoencoder input and output for different bottleneck dimensions. We can observe that the error reaches a plateau for a bottleneck dimension of $N=3$. We thus consider here that the pertinent number of BAD parameters to study the structure in our system is therefore three.

\begin{figure}[h!]
\centering
\includegraphics[width=8cm]{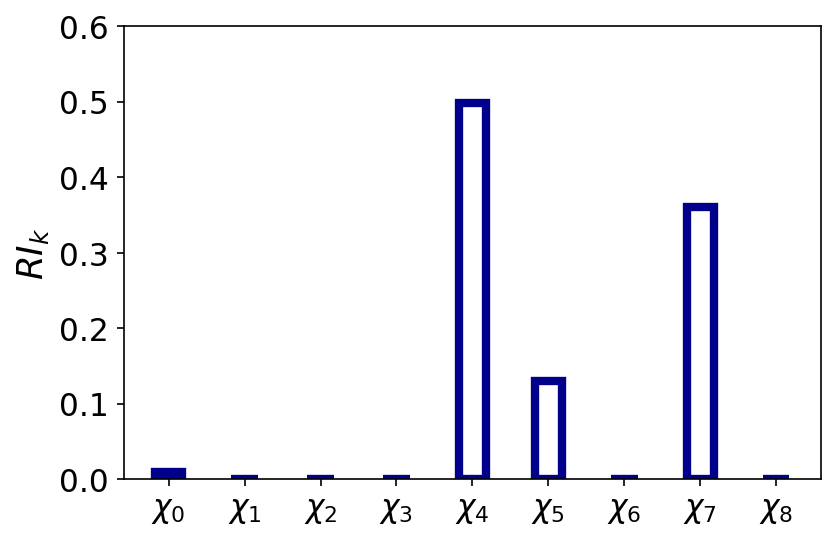}
\caption{Relative importance index of the BAD parameters obtained from the input perturbation method. The index is calculated by perturbating consecutively each input parameter and measure how much this perturbation affected the output. The parameters $\{ \chi_4,\chi_5,\chi_7\}$ are shown to be the most relevant ones to study the different structures in the system.}
\label{fig:RIindex}
\end{figure}

Our next step is to determine which are the three most pertinent BAD parameters out of nine we computed for each atom. To this end, we calculate the relative importance (RI) of each parameter applying the perturbation method described in Section \ref{subsubsec:RelevantParameters}. The results are shown in Fig \ref{fig:RIindex}. From this figure, we can clearly observe the most significant three parameters: the triplet $\{ \chi_4,\chi_5,\chi_7\}$. From here on in this study, we will only focus on these three BAD parameters to study the substructures within the studied Ni nano-object.

\subsection{Clustering methods combination} 
\label{subsec:CompareClustering}

We determined that the most pertinent BAD parameters to study the emerging substructures in our numerical sample are $\{ \chi_4,\chi_5,\chi_7\}$. Here, we apply two clustering algorithms on these parameters, K-means and DBSCAN, and then we combine the outcome of the obtained clusters. The dataset used for the clustering processes is the same that we used for extract BAD parameters using the autoencoder (we also verified that using different atomistic configurations extracted after the onset of plasticity gives similar results) .

\begin{figure}[h!]
\centering
\includegraphics[width=10cm]{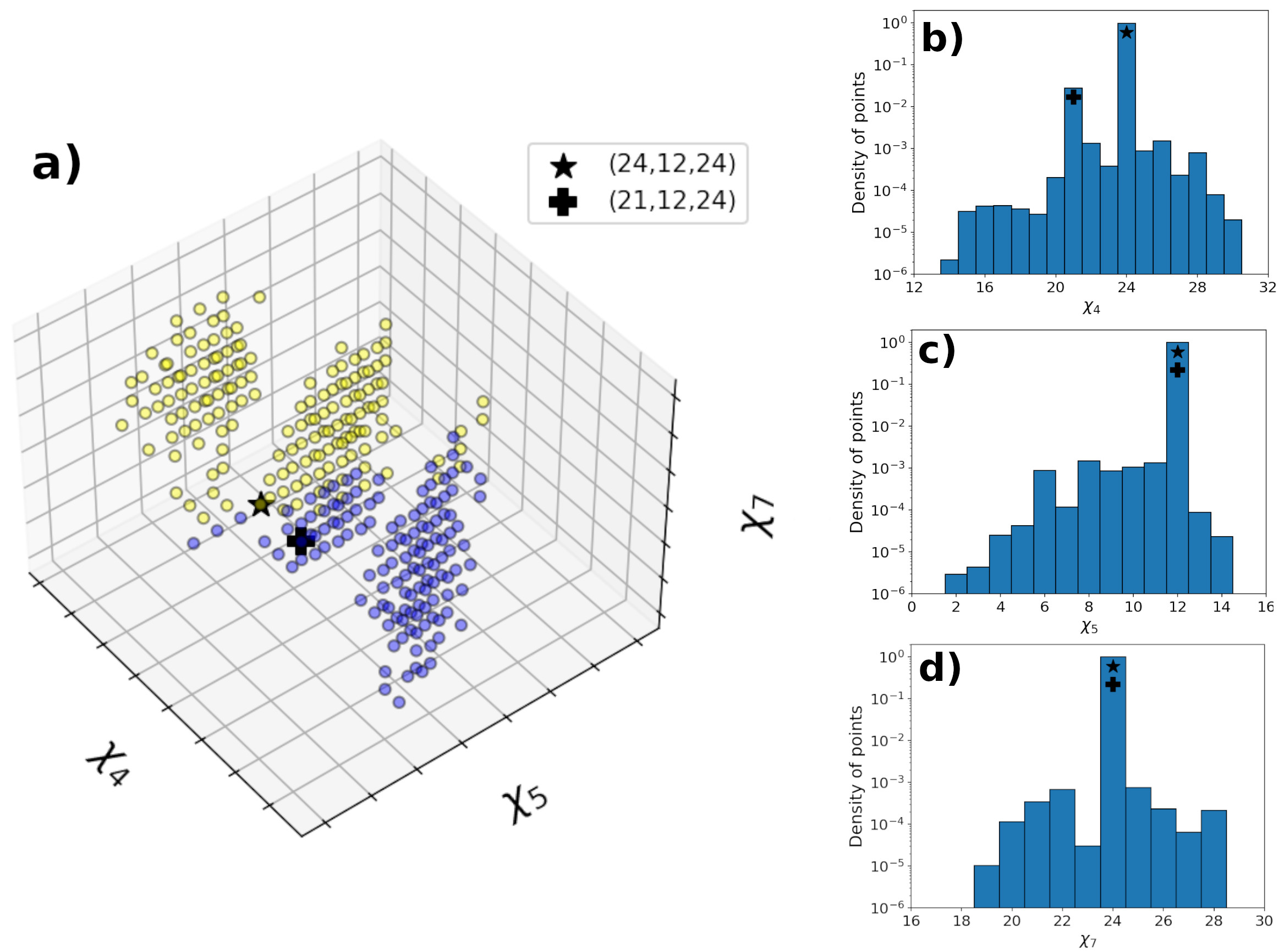}
\caption{The upper part of the figure (a) represents a Scatter plot of the BAD parameters projected on the most relevant parameters: $\{ \chi_4,\chi_5,\chi_7\}$. The colors correspond to the substructures detected with K-means clustering.
The lower part shows histograms showing the distribution of $\chi_4$ (b), $\chi_5$ (c), $\chi_7$ (d). On the $\chi_4$ histogram, we observe that most of the atoms are concentrated on two values, while on the $\chi_5$  and $\chi_7$ histograms, most atoms are concentrated on a single value. Note that two combinations of $\{ \chi_4,\chi_5,\chi_7\}$, $\{ 24,12,24\}$ and $\{ 21,12,24\}$, marked with a star and a plus symbol, respectively, contain most of the atoms and correspond in practice to the FCC and the stacking fault (HCP) structures. The K-means clustering method aims here at separating these two points corresponding to two very close clusters.}
\label{fig:histograms}
\end{figure}

On Fig \ref{fig:histograms} (a), we show a scatter plot representing the atoms spatial distribution in the $\{ \chi_4,\chi_5,\chi_7\}$ space, as well as the result of the K-means clustering.  Note that as the BAD parameters are integers, one position in the $\{ \chi_4,\chi_5,\chi_7\}$ space can correspond to many atoms. Each point represented here corresponds to at least one atom. On this figure, we can see that the algorithm detected two clusters: one in dark blue and the other in yellow.

\begin{figure}[h!]
\centering
\includegraphics[width=13.7cm]{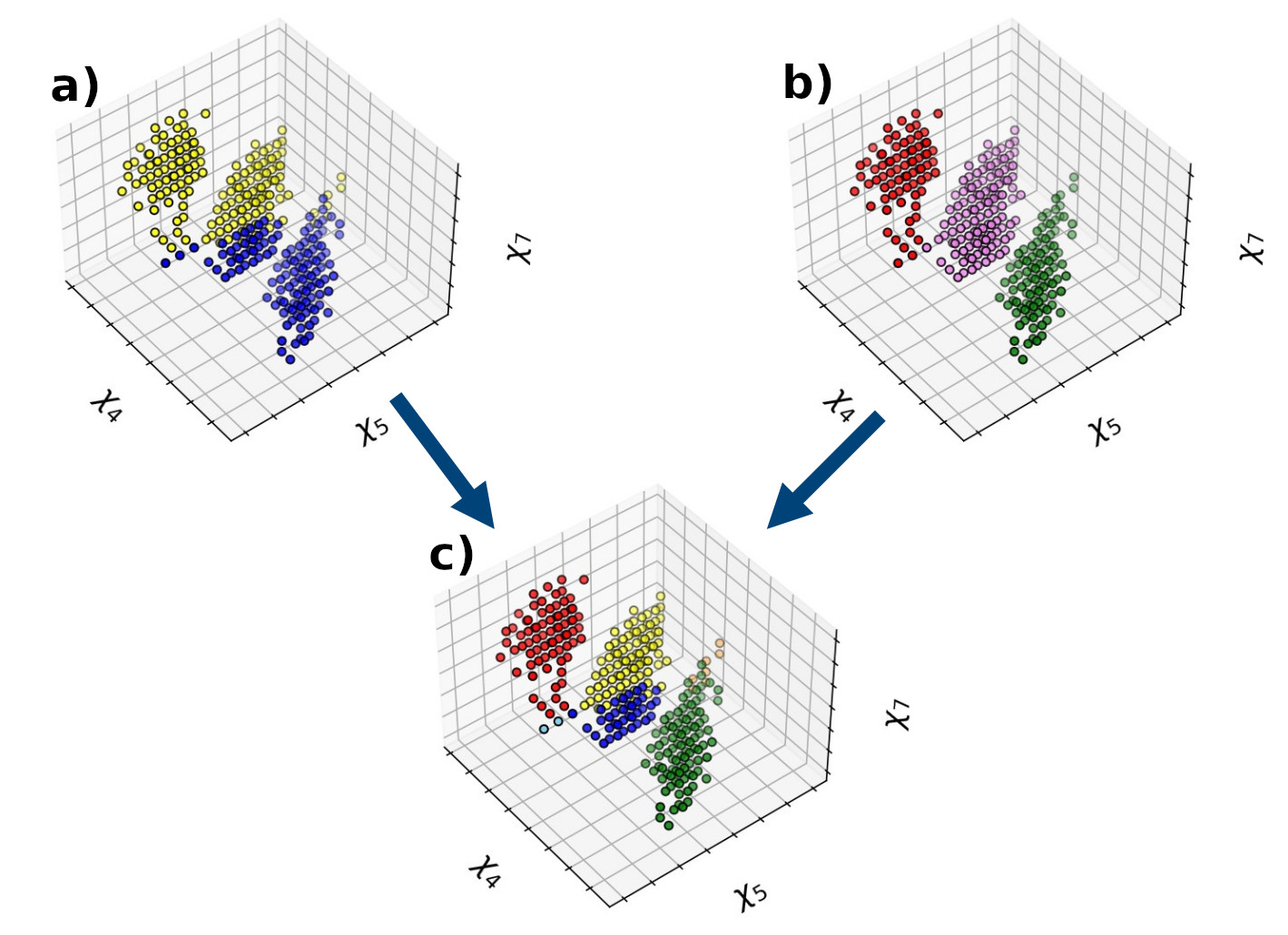}
\caption{Scatter plots of the BAD parameters projected on the most relevant parameters: $\{ \chi_4,\chi_5,\chi_7\}$. The colors correspond to the clustering methods used to detect different structures with: (a) K-means clustering, (b) DB scan clustering combined with logistic regression classification, and (c) a combination of the clusters obtained through (a) and (b). Each point corresponds to at least one atom, the distribution for each parameter can be seen on Fig \ref{fig:histograms}. In total, six differents structures were detected through this combination of two clustering methods: FCC (yellow), HCP in the stacking fault (blue), upper and lower part of the dislocation lines (green and red), the interaction between dislocation line and stacking fault (skyblue) and the interaction between two dislocation lines (orange).}
\label{fig:scatterBAD}
\end{figure}

As detailed in Section \ref{paragraph:Kmeans}, the aim of using the K-means clustering is to detect the main crystalline structures, located close to each other in the BAD parameter space \citep{ackland_applications_2006}.
While the Fig \ref{fig:histograms} (a) clearly shows three blocks, the two detected clusters seems to correspond to half of the central block.
To understand what was detected by the K-means algorithm, we look on Fig \ref{fig:histograms} the distribution of $\chi_4$ (b), $\chi_5$ (c) and $\chi_7$ (d). From this figure, we can see that the large majority of the atoms are concentrated on two close positions in the $\{ \chi_4,\chi_5,\chi_7\}$ space: $\{24,12,24\}$ and $\{21,12,24\}$ associated with a star and a plus symbol, respectively, in Fig \ref{fig:histograms} (a). The points surrounding the two main substructures are due to thermal fluctuations and deformation imposed by mechanical loading. On \citep{ackland_applications_2006}, these two configurations were defined as corresponding to FCC and HCP, respectively. As HCP corresponds to the stacking fault structure in a FCC crystal \citep{hull_chapter_2011}, we can expect that most of the atoms will be in these two structures. 
From what we can see on Fig \ref{fig:scatterBAD} (a), K-means seems to have succeeded in separating these two configurations located within the central block. This will be confirmed in the next part focusing on analysing the structures detected by the clustering algorithms.

While the K-means algorithm successfully separated the two locations in the $\{ \chi_4,\chi_5,\chi_7\}$ space, we can see on Fig \ref{fig:histograms} (a) that it fails in identifying two clusters located further away from the central block, corresponding to different substructures. To detect them, we use another clustering method: the DBSCAN combined with the logistic regression classifier detailed in \ref{paragraph:DBScan} and \ref{paragraph:LogisticRegression}, respectively. We show its results on Fig \ref{fig:scatterBAD} (b). From this figure, we can see that the DBSCAN successfully managed to separate into three clusters the three blocks observed in the scatter plot. After measuring the proportion of data points contained in each cluster, it appears that the central cluster (in purple) contains $99.8\%$ of the points, while the two remaining clusters contains $0.1\%$ each. This confirms that most of the atoms are located in the main structures, corresponding to the central block, making it difficult for the K-means clustering to detect the two smaller blocks.

Finally, to detect the main crystalline structures while being able to discern other structures containing fewer atoms in the system, we combine the results obtained by the two clustering methods (K-means shown on Fig \ref{fig:scatterBAD} (a), DBSCAN shown on Fig \ref{fig:scatterBAD} (b)) as highlighted on Fig \ref{fig:scatterBAD} (c). From this figure, we can see that we managed to detect a total of six different clusters: the dark blue and yellow ones from the central structures from the central block, containing the $99.8\%$ of the atoms, corresponding to the main crystalline structures. The red and green cluster, corresponding to the two other aggregation of points clearly visible in the scatter plot of Fig \ref{fig:scatterBAD} (c). 
Finally, the sky blue and the orange clusters, which only corresponds to few atoms, and detected only in few time steps.

\subsection{Analysis of detected structures} 
\label{subsec:DBScanclustering}

On the previous section, by combining two clustering methods, we managed to extract six different clusters from our dataset by a projection in the $\{ \chi_4,\chi_5,\chi_7\}$ space. We now focus on analyzing the structures associated with each cluster. 
To validate our model, our test dataset consists in different time steps of the same simulation than the one used for training the model, as well as to different simulations performed with a different initial MD velocity random seed. Thus, the defects in the test dataset are at different locations compared to those in the training dataset.

\begin{figure}[h!]
\centering
\includegraphics[width=13cm]{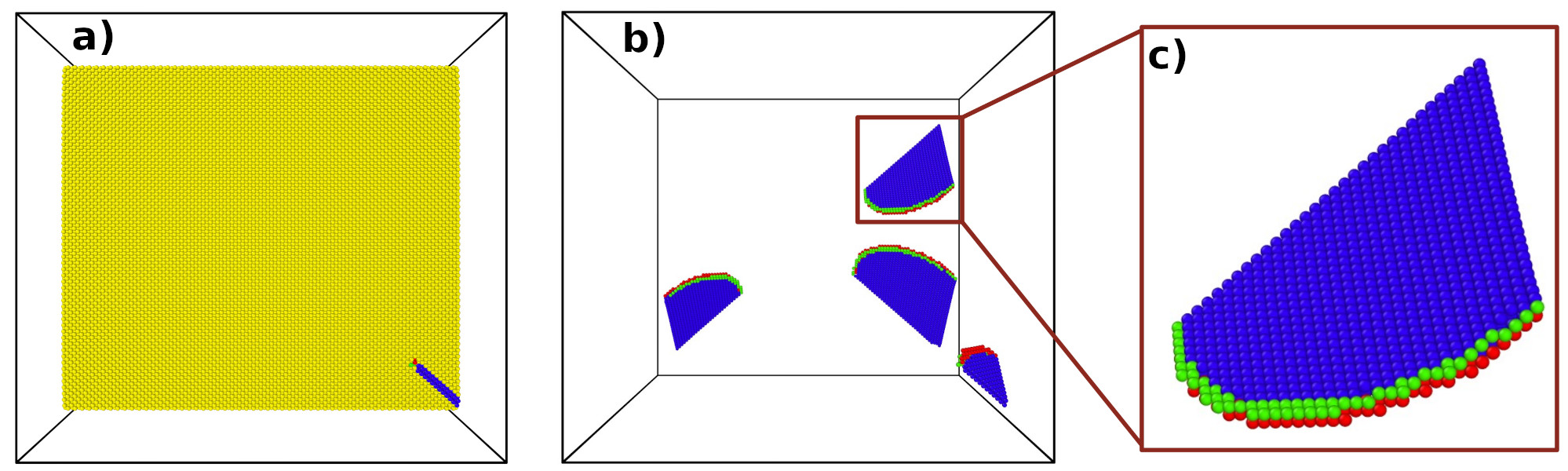}
\caption{Snapshot of the system. The atoms are colored depending on their cluster as shown in Fig \ref{fig:scatterBAD} c). On panel (a), we observe that most of the atoms belongs to the yellow cluster which we associate with the FCC structure. On (b), we show the same snapshot after filtering out the FCC structure. We can observe that the combination of the two clustering methods is capable to separate the atoms linked with dislocation lines (green and red atoms) and those correlated with stacking faults (dark blue atoms).}
\label{fig:snapshots}
\end{figure}

We show on Fig \ref{fig:snapshots} (a) a snapshot made with OVITO \citep{stukowski_visualization_2009} of the Ni nano-cube deformed plastically, with an applied strain of $\Delta \epsilon=0.02\%$ beyond the yield point, for data extracted from our test dataset.  On this snapshot, the atoms are colored corresponding to their clusters . For instance, the atoms contained in the yellow cluster on Fig \ref{fig:scatterBAD} (c) will be colored in yellow on Fig \ref{fig:snapshots}.
On Fig \ref{fig:snapshots} (a), we can see that most of the atoms are contained in the yellow cluster which we can then associate with the FCC structure as our system is a Ni FCC nano-crystal.

After filtering out the atoms in the FCC structure, we can now see on Fig \ref{fig:snapshots} (b) atoms in three different structures. Fig \ref{fig:snapshots} (c) corresponds to a zoom within Fig \ref{fig:snapshots} (b) to highlight more clearly the detected substructures. 
From this figure, the atoms colored in green and red follows what looks like a dislocation line, while the atoms in dark blue seems to correspond to the stacking fault created after the glide of a partial dislocation. By using the algorithm DXA \citep{stukowski_automated_2012} implemented in OVITO \citep{stukowski_visualization_2009}, we could confirm our prediction: the green and the red atoms are surrounding a partial dislocation line. The relative position of the green and red atoms depends on the Burgers vector as well as on the direction of the dislocation line (see the appendix \ref{sec:Annexe:StructureAndDislocationLines} for more details). From this, we can deduce that the dark blue atoms correspond to HCP stacking fault structures generated by the gliding of a partial dislocations (as mentioned above in FCC crystals stacking faults have a HCP structure).

The two remaining clusters detected in in Fig \ref{fig:scatterBAD} c), the orange and the sky blue one, are very rarely observed during the time steps recorded during the deformation. The orange structure is observed when two dislocation lines are located close to each other while the sky blue one is detected during the interaction between a dislocation line and a stacking fault, as shown later.

\begin{figure}[h!]
\centering
\includegraphics[width=13.5cm]{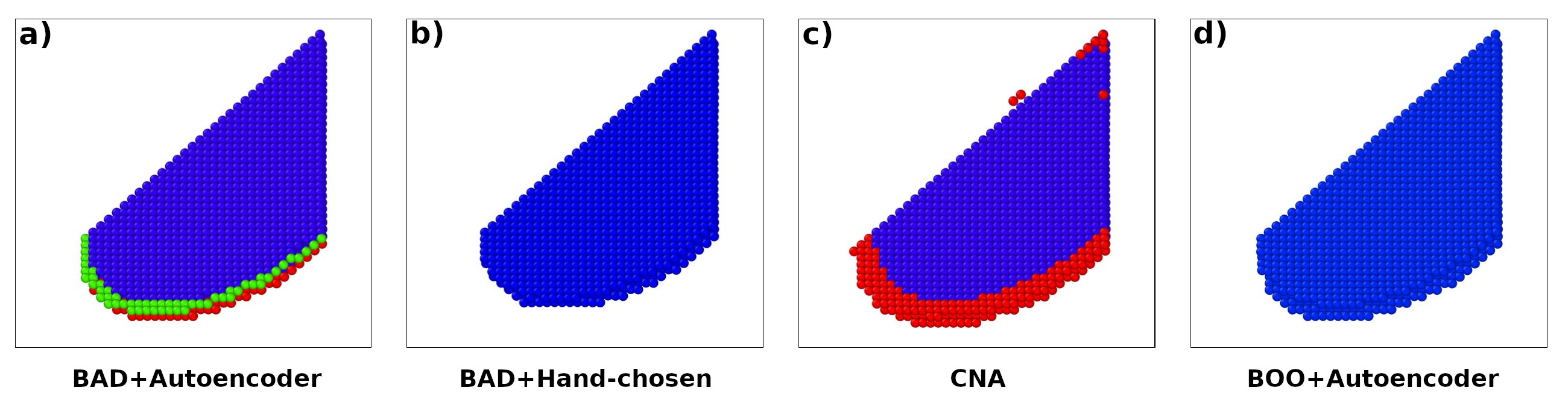}
\caption{Snapshot of the system. The atoms are colored depending on their surrounding structure obtained from (a) our combination of BAD and autoencoder, (b) hand-chosen criteria applied on BAD from \citep{ackland_applications_2006}, (c) the CNA method \citep{honeycutt_molecular_1987} from OVITO \citep{stukowski_visualization_2009}, and (d) the combination of BOO and autoencoder as detailed in \citep{boattini_unsupervised_2019}. On (a), the blue atoms are associated with the stacking fault while the atoms in red and green are associated with the dislocation lines, on (b) the atoms in blue correspond to an HCP structure, on (c) the atoms in blue correspond to an HCP structure, and those in red are labeled "other" by the algorithm, and finally on (d) the atoms in blue are associated with the stacking fault.
Note that for the four structure detection methods shown here, we removed the surface atoms beforehand.}
\label{fig:snapshotsCompare}
\end{figure}

On Fig \ref{fig:snapshotsCompare}, we compare our structure detection method, shown in panel (a), with three other existing methods: (i) hand-chosen criterion using the BAD parameters from \citep{ackland_applications_2006} (Fig \ref{fig:snapshotsCompare} (b)), (ii) the CNA method \citep{honeycutt_molecular_1987,stukowski_structure_2012} implemented in OVITO ,which is one of the most commonly used method for structure detection (Fig \ref{fig:snapshotsCompare} (c)) and which consists of using hand-chosen criterion on a radial-distribution based local order parameter, and (iii) the autoencoder approach using the BOO parameters from \citep{boattini_unsupervised_2019} which served as an inspiration for this paper (Fig \ref{fig:snapshotsCompare} (d)). For this comparison, we filtered beforehand the surface atoms following the method explained in section \ref{subsec:UnsupervisedLearning}. 

\begin{figure}[h!]
\centering
\includegraphics[width=13cm]{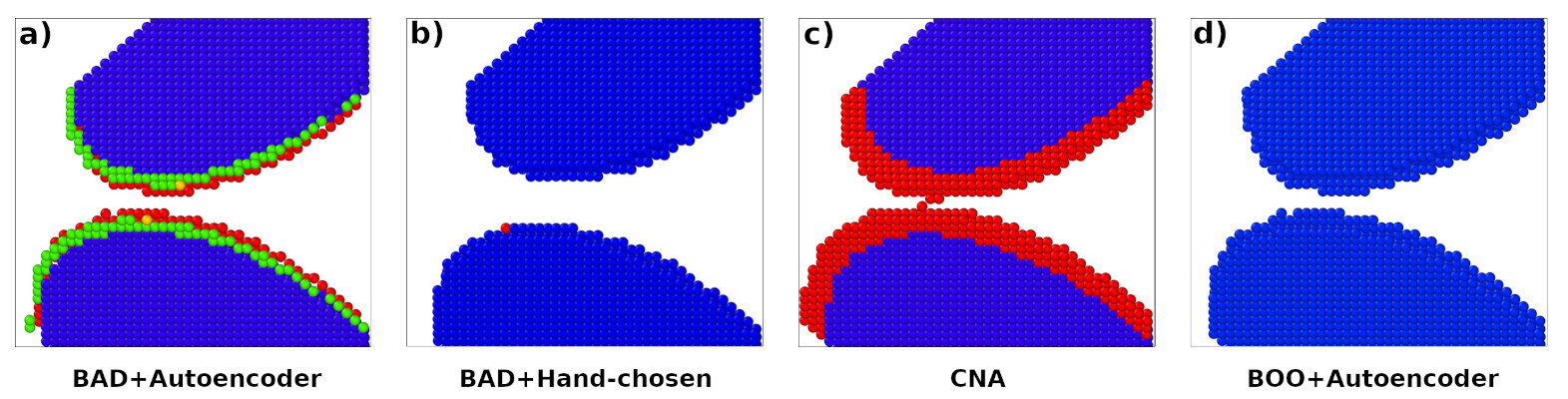}
\caption{Snapshot of two dislocations about to interact. These snapshots were obtained after filtering out the FCC structure. The atoms are colored depending on their surrounding structure obtained from (a) the combination of BAD and autoencoder, (b) hand-chosen criteria applied on BAD from \citep{ackland_applications_2006}, (c) the CNA method \citep{honeycutt_molecular_1987} from OVITO \citep{stukowski_visualization_2009}, and (d) the combination of BOO and autoencoder as detailed in \citep{boattini_unsupervised_2019}. We can remark on (a) two atoms belonging to the orange cluster from Fig \ref{fig:scatterBAD} (c) within the dislocation lines. Atoms from this cluster were observed when two dislocations are close to each other. We can also remark on (b) one atom in red corresponding to the unassigned category on the hand-chosen criteria applied to BAD from \citep{ackland_applications_2006}.}
\label{fig:Annexe:SnapshotsCompare2}
\end{figure}

\begin{figure}[h!]
\centering
\includegraphics[width=13cm]{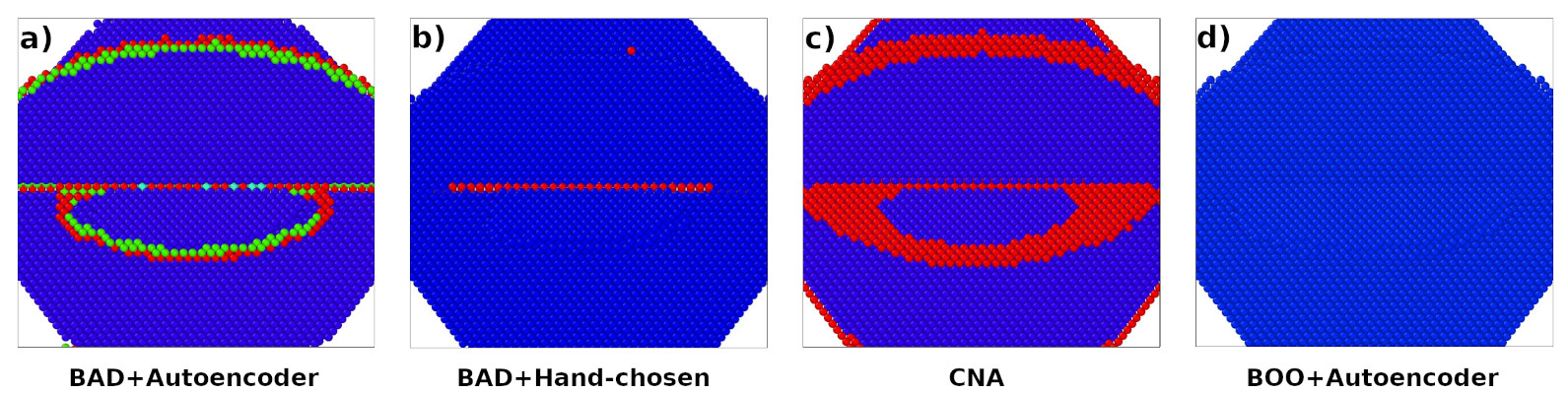}
\caption{Snapshot of the system focusing on the interaction between two dislocations, one of them crossing the stacking fault of the other. These snapshots were obtained after filtering out the FCC structure. The atoms are colored depending on their surrounding structure obtained from (a) the combination of BAD and autoencoder, (b) hand-chosen criteria applied on BAD from \citep{ackland_applications_2006}, (c) the CNA method \citep{honeycutt_molecular_1987} from OVITO \citep{stukowski_visualization_2009}, and (d) the combination of BOO and autoencoder as detailed in \citep{boattini_unsupervised_2019}. We can observe on (a) atoms belonging to the sky blue cluster from Fig \ref{fig:scatterBAD} (c). Atoms from this cluster are observed after the interaction between a dislocation line and a staking fault. }
\label{fig:SnapshotsCompare3}
\end{figure}

On Fig \ref{fig:snapshotsCompare}, we only focus on one dislocation line and the associated stacking fault. From this figure, we can see that our method is able to capture with much more details the different structures present within a FCC crystal plastically deformed. To further emphasize the efficiency of our method,  on Fig \ref{fig:Annexe:SnapshotsCompare2} we show a snapshot of the system with two dislocation lines about to interact and on Fig \ref{fig:SnapshotsCompare3} we show the interaction between a dislocation line and a stacking fault.

Out of the three methods used for comparison, two were mostly designed to detect the main crystalline structures (FCC, HCP or BCC): the hand-chosen criterion using the BAD parameters and CNA.
We can still remark that CNA, which also relies on hand-chosen criterion, is able to capture the location of the dislocation line but with less accuracy than our method: the number of atoms close to the line defect is much larger than in our method. Also, this method is not able to capture sub-structures associated with plasticity. The atoms outside of the main crystalline structures, such as those in the line defect, are labelled as "other" (in red in Fig \ref{fig:snapshotsCompare} (c)). 
 The hand chosen method using the BAD also categorizes from time to time atoms in the category "other", like in Fig \ref{fig:Annexe:SnapshotsCompare2} (b), but it is not able to capture the line defect.

The last method, the autoencoder applied to the BOO parameters, is also designed to detect automatically the different structures present within the system. However, when applying the method to our system, only two structures were detected: one associated with FCC and the other associated with HCP (the stacking fault). Also, only two structures were visible on the lower dimensional subspace over which the clustering was applied when following the method (see appendix \ref{sec:Annexe:ClusteringBOOP}). We thus conclude that the BOO parameters are not the most pertinent parameters to detect the structures within a plastically deformed crystal.

%
%On Fig \ref{fig:Annexe:SnapshotsCompare3}, we present another snapshot of the system focusing on the interaction between two dislocations, one of them crossing the stacking fault of the other. These snapshots were obtained after filtering out the FCC structure. We show on (a) the structures detected by our method. On top of the stacking fault (in dark blue) an the dislocation line (green and red), we also observe atom in sky blue at the intersection between the stacking fault and the dislocation line. This correspond to another cluster from our method which was also observed for a handful of atoms. On (b), we observe that the atom at the intersection were considered as "unassigned" by the method using hand-chosen criteria applied to BAD \citep{ackland_applications_2006}. 
%
%From Fig \ref{fig:Annexe:SnapshotsCompare2} and Fig \ref{fig:Annexe:SnapshotsCompare3}, we can see that our method offer a better detection of the structures following the dislocation lines.

Overall, these comparisons show that our method is able to detect more substructures related to early plasticity than the other presented approaches. The method presented in this paper is also able to capture the contours of the dislocation lines that interact with a better precision than CNA as we can see on Fig \ref{fig:SnapshotsCompare3}, where in panel (a) we can clearly distinguish  the profile of two interacting dislocation lines (see Fig. \ref{fig:Annexe:ComposeDisloInteraction} (a) of \ref{sec:Annexe:InteractionDislo} for a wider representation and a comparison with DXA).

\vspace{5mm}

In this paper, we focused our study on substructures associated to plastic deformation up to $10\%$: the dislocation lines, their interactions and the stacking fault. By applying our method to a crystal deformed up to an applied strain of $40\%$, one more cluster was detected. However, we could not relate this cluster to particular structure. As the corresponding atoms where located in very deformed regions at the intersections of many dislocation lines, it is likely that these atoms are located in a so deformed regions that no crystalline substructure can be associated and are in an almost amorphous configuration. These results of the method are shown on \ref{sec:Annexe:HigherDeformation} of the supplementary materials.

Also, one room of improvement for our method is the definition of the BAD parameters. Indeed, in this paper, we used the original definition of the BAD parameters in which the bond-angles are separated in hand-chosen ranges. While the used BAD parameters, coupled with unsupervised learning, allow to detect more substructures than the other presented approaches, it is not optimized for all the crystalline systems. Modified BAD parameters exist but consist of adapting the hand-chosen ranges to a specific material such as a specific crystalline alloy \citep{amodeo_atomistic_2014}. Thus, one way to achieve a better detection of substructures associated with plastic deformation would be to improve of the BAD approach by creating a more universal BAD-based parameter, i.e., not relying on hand-chosen criteria. It is however outside the scope of the present study which principally aims at introducing the unsupervised learning approach.

\section{Conclusion}
\label{sec:Conclusion}

In conclusion, we improved a method initially developed to detect automatically structures in colloidal materials. Through this method, we were able to finely detect different structures present in a Ni FCC crystal under plastic deformation.
This algorithm uses the BAD parameters to describe the local environment of the atoms. Thanks to an autoencoder associated with a perturbation method, the most relevant BAD parameters were extracted. Then, by applying on these selected parameters two clustering methods, Kmeans and DBscan as well as the logistic regression classifier, different local structures have been successfully extracted. This procedure was able to detect the local structures present within the dislocation lines with a higher degree of precision compared to the commonly used CNA method. Our procedure is also more efficient than a method relying on hand-chosen criteria applied to BAD parameters or than combining autoencoder with BOO parameters.

Overall, this study shows that unsupervised learning applied to the BAD parameters is a promising approach to obtain more precise structure detection within crystalline materials. As the method detailed in this paper is based on parameters using hand-chosen ranges in its definition, it is not directly applicable to other crystalline system, especially alloys. While modified BAD parameters adapted for specific systems exist \citep{amodeo_atomistic_2014}, developing a method to automatically find the optimal ranges to define the BAD parameters for any materials would allow to achieve an universal method for structural detection for crystalline materials.

\section*{Declaration of competing interest}
\label{sec:DeclarationInterest}
The authors declare that they have no known competing financial interests or personal relationships that could have appeared to influence the work reported in this paper.

\section*{Data availability}
\label{sec:Datavailability}
The data and scripts used for the paper are available on open-access on Zenodo \citep{barbot_armand_2023_7582668}.

\newpage

\appendix

\section{Linking detected structures and dislocation lines}
\label{sec:Annexe:StructureAndDislocationLines}

\begin{figure}[h!]
\centering
\includegraphics[width=12cm]{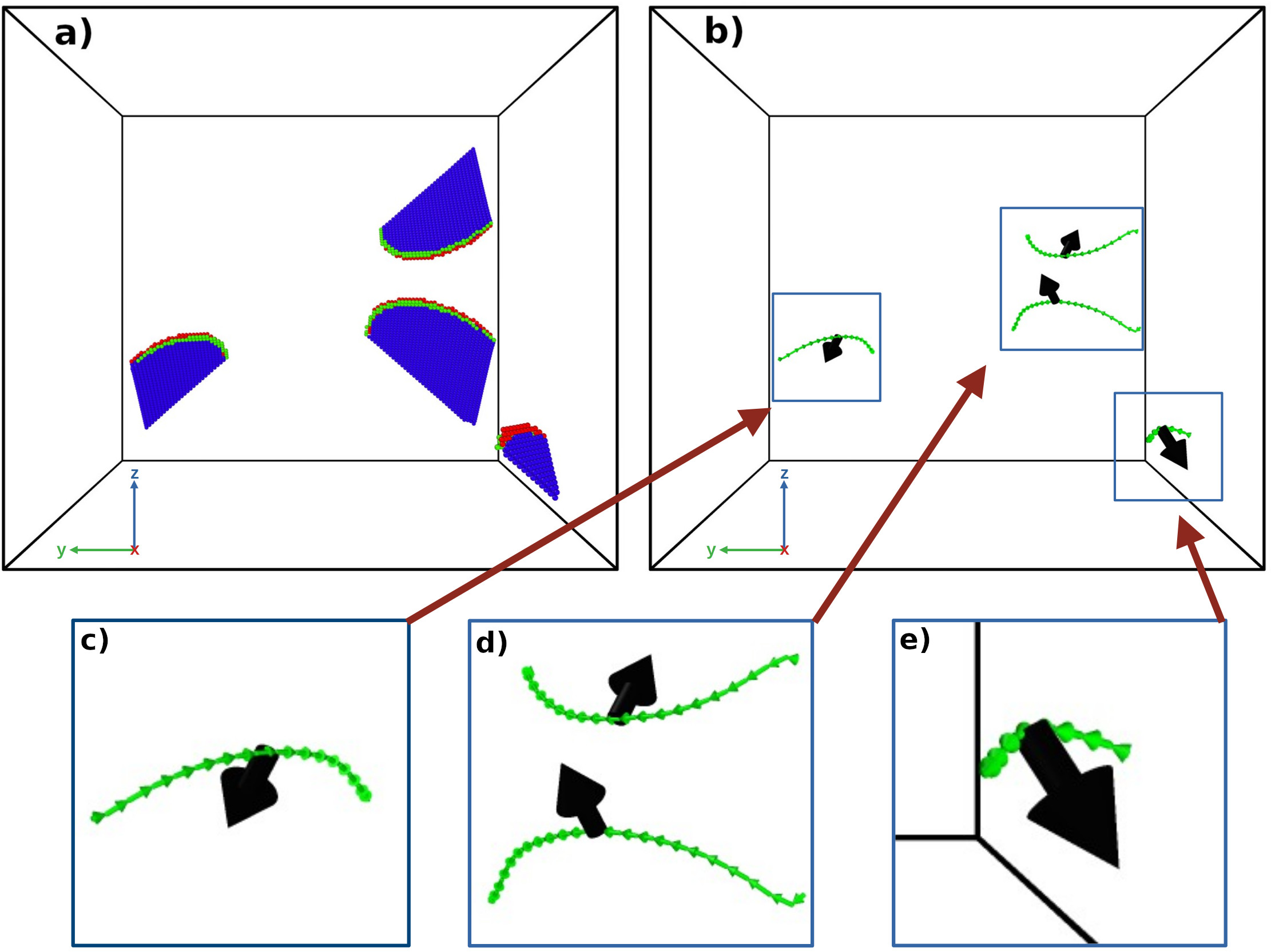}
\caption{Snapshot of the system. On (a) the FCC atoms are filtered out to only keep the atoms in a stacking fault (in blue) and in a dislocation line (in green and in red). To understand the relative position between the green and the red atoms, we show on (b) the dislocation lines (in green) obtained from DXA \citep{stukowski_structure_2012}. The green arrows on the dislocation lines corresponds orientation of the dislocation line. The black arrows corresponds to the Burgers vectors direction. The figures (c), (d) and (e) corresponds to zooms on the dislocation lines for easier visualization. We remark that when the dislocation line direction goes towards the trigonometric direction around the z-axis while the dislocation goes toward the center of the dislocation loop (as in (c) and (e)), the red atoms are above the green ones. }
\label{fig:Annexe:SnapshotsDislocationDirection}
\end{figure}

On Fig \ref{fig:Annexe:SnapshotsDislocationDirection} we show a comparison between the structures detected with our method (a) and the dislocation lines obtained from DXA \citep{stukowski_structure_2012} (b). On Fig \ref{fig:Annexe:SnapshotsDislocationDirection} (b) the black arrows correspond to the Burgers vector of each dislocation and the green arrows show the dislocation lines directions.

\section{Autoencoder and clustering applied to BOO parameters}
\label{sec:Annexe:ClusteringBOOP}

To compare more precisely the BOO and BAD parameters to detect local structure, we reproduced the method from \citep{boattini_unsupervised_2019}, which served as an inspiration for our method, and applied it to our data set. The method consists basically on calculation the BOO parameters for each atoms and training the autoencoder to find the optimal bottleneck dimension. Then, our method diverges from ours as they apply gaussian clustering directly to the output of the bottleneck. 

\begin{figure}[h!]
\centering
\includegraphics[width=8cm]{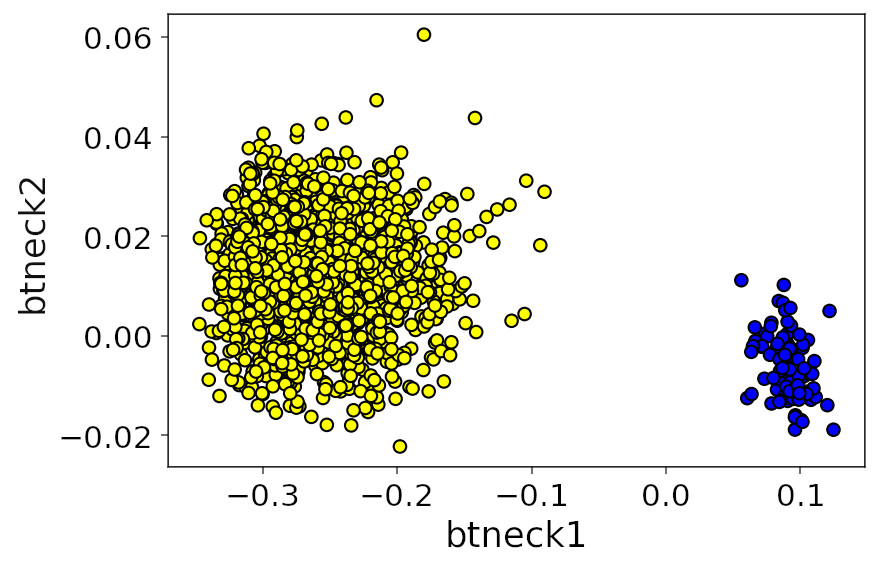}
\caption{Scatter plot showing the output of the 2-dimensional bottleneck obtained by applying the method from \citep{boattini_unsupervised_2019} on our training data set. We calculate for each atom their BOO parameters using pyscal \citep{menon_pyscal_2019}. We then obtained an optimal dimension of two at the bottleneck. Finally, we applied gaussian clustering at the bottleneck output detected two clusters: the yellow one corresponding to FCC structure and the blue one corresponding to HCP structure. As no other clusters can be seen, we observe that BOO is not able to separate the dislocation lines from the staking faults.}
\label{fig:Annexe:BOOClusters}
\end{figure}

By applying this approach, we obtain an optimal bottleneck dimension equal to two. We then show on Fig \ref{fig:Annexe:BOOClusters} a scatter plot showing for each atom their position in the space of the two bottleneck dimensions that we name $\{btneck1,btneck2\}$. By following the clustering method detailed in \citep{boattini_unsupervised_2019}, we extract two clusters: one in yellow corresponding to the FCC structure and one in blue corresponding to the HCP structure (in the stacking faults). We can see that no other clusters that could correspond to the structures within the dislocation lines are visible on Fig \ref{fig:Annexe:BOOClusters}. From this, we deduce that the BOO parameter is not pertinent to separate the dislocation lines from the other structures in a crystal under plastic deformation.

\section{Interaction between dislocations}
\label{sec:Annexe:InteractionDislo}

\begin{figure}[h!]
\centering
\includegraphics[width=9cm]{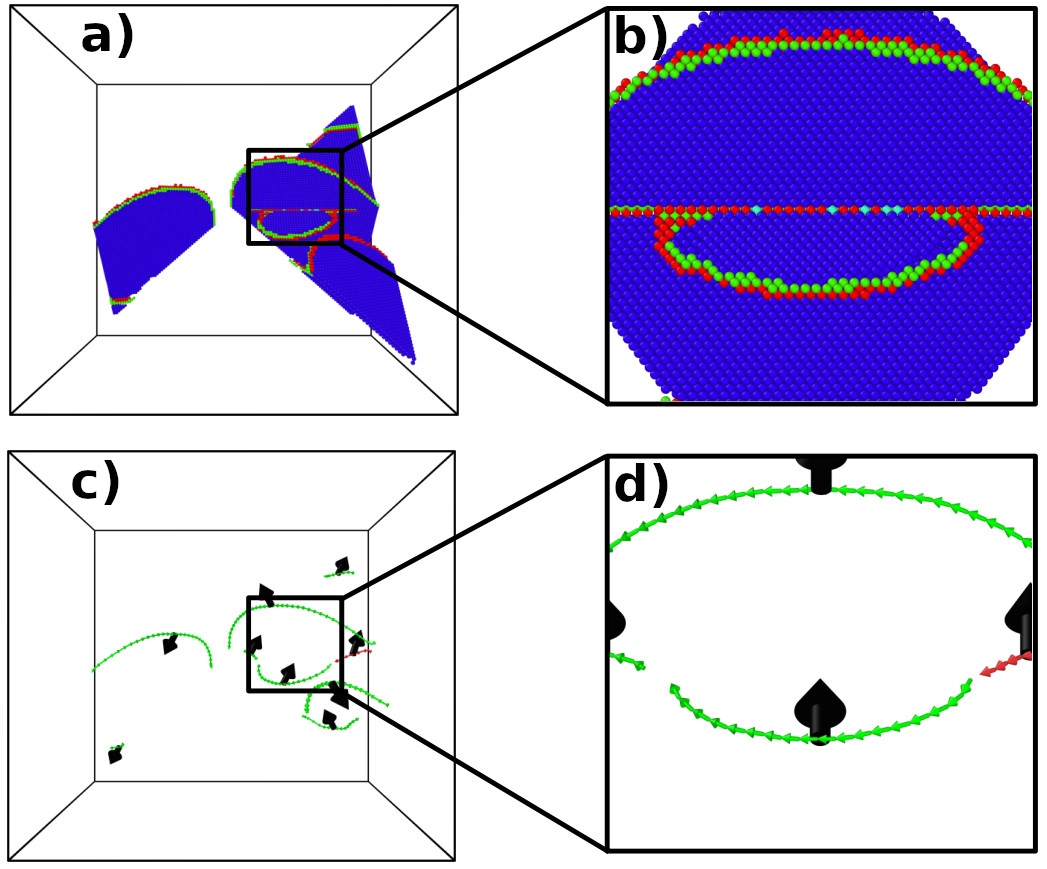}
\caption{Snapshot of two interacting dislocation. The detection and the shape of the dislocation lines obtained using the unsupervised learning approach in (a) and (b) match the one obtained by the DXA algorithm from OVITO as shown in (c) and (d). }
\label{fig:Annexe:ComposeDisloInteraction}
\end{figure}

As shown in Fig. \ref{fig:Annexe:ComposeDisloInteraction} the unsupervised learning approach is also able to capture the interaction between dislocations. Fig. \ref{fig:Annexe:ComposeDisloInteraction} (a) shows a wide view of two interacting dislocations and (b) focuses more specifically on the interaction region. The Fig. \ref{fig:Annexe:ComposeDisloInteraction} (c) and (d) show for the same viewpoints as (a) and (b), respectively, the location of the dislocation lines obtain by the DXA algorithm implemented in OVITO.  We can clearly observe that the position of the dislocation lines predicted by the unsupervised learning approach of this manuscript matches the one predicted by the analysis made with OVITO.

\section{Application to more deformed systems}
\label{sec:Annexe:HigherDeformation}

While our study mostly focus on detecting the substructures associated with early plasticity, we detail the results obtained by applying our method to a Ni crystal deformed up to $40\%$ with uniaxial compression, having a much higher dislocation density (about $10^{17}$ $m^{-2}$) than the systems deformed up to $10\%$ (about $10^{15}$ $m^{-2}$). 

\begin{figure}[h!]
\centering
\includegraphics[width=5.5cm]{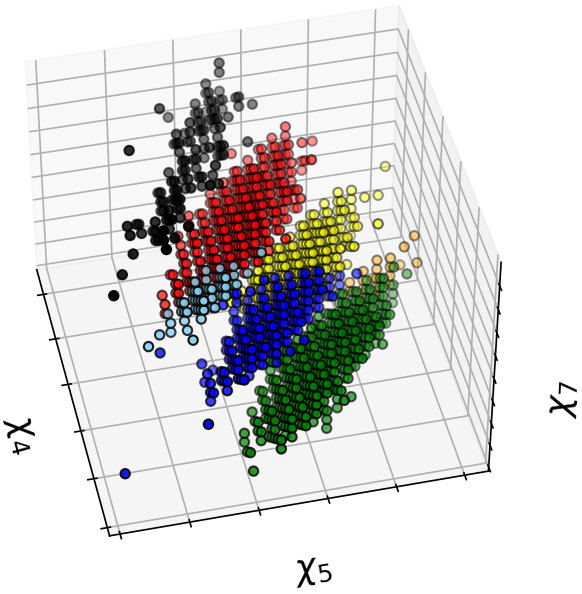}
\caption{Scatter plot of the BAD parameters projected on the most relevant parameters: $\{ \chi_4,\chi_5,\chi_7\}$ for a system deformed with an applied strain of 0.4. At this strain level, one more cluster (in black) was detected on top of the six previously detected for the less deformed system. While it is difficult to relate this cluster to any specific structure, we believe that it corresponds to regions so highly deformed locally that no crystalline substructure can be associated to the atoms in this cluster.}
\label{fig:Annexe:BADHighDeformClusters}
\end{figure}

By applying our method, we obtained again that the most pertinent parameters are $\{ \chi_4,\chi_5,\chi_7\}$. On Fig \ref{fig:Annexe:BADHighDeformClusters} we show a scatter plot of the most pertinent BAD parameters for the atoms of Ni crystal deformed up to $40\%$. With respect to the clusters shown in Fig. \ref{fig:scatterBAD}, one more cluster (in black) was detected.

\begin{figure}[h!]
\centering
\includegraphics[width=13cm]{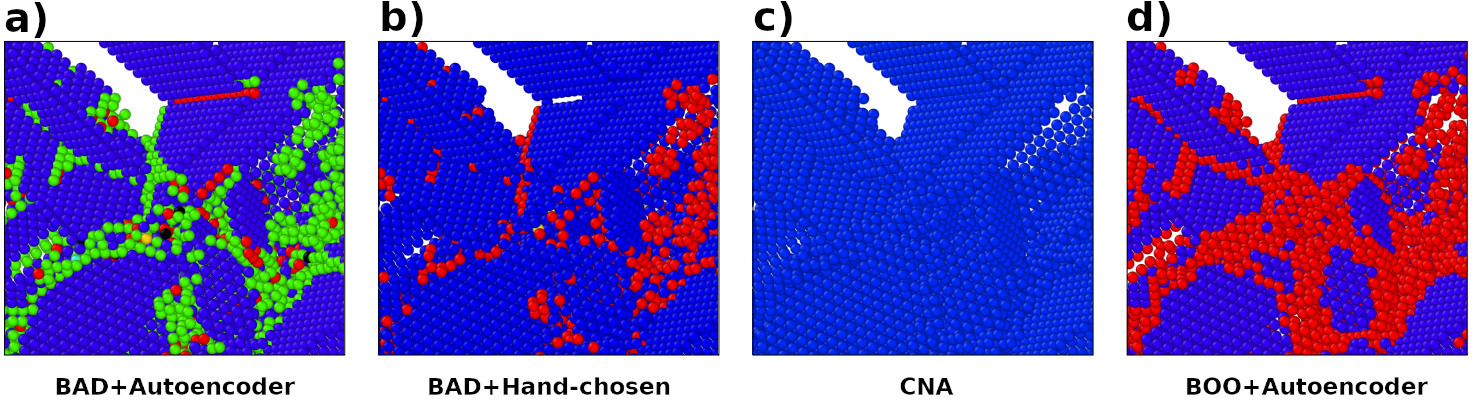}
\caption{Snapshots of the system focusing on a location with atoms from the black cluster of Fig \ref{fig:Annexe:BADHighDeformClusters}. These snapshots were obtained after filtering out the FCC structure. The atoms are colored depending on their surrounding structure obtained from (a) the combination of BAD and autoencoder, (b) hand-chosen criteria applied on BAD from \citep{ackland_applications_2006}, (c) the CNA method \citep{honeycutt_molecular_1987} from OVITO \citep{stukowski_visualization_2009}, and (d) the combination of BOO and autoencoder as detailed in \citep{boattini_unsupervised_2019}. We can observe on (a) atoms belonging to the black cluster. Atoms from this cluster are observed at locations where many dislocation lines are intertwined. We can also remark that on (b), one atom is colored in yellow. It means that the used algorithm detected it as being in a BCC structure.}
\label{fig:Annexe:SnapshotsCompare4}
\end{figure}

To determine to which substructure is associated the black cluster, we show on Fig \ref{fig:Annexe:SnapshotsCompare4} (a) a snapshot focusing on a location with atoms from the black cluster of Fig \ref{fig:Annexe:BADHighDeformClusters}. These atoms are located on regions where many dislocation lines are intertwined. However only a handful of atoms are associated with this cluster, and it is difficult to associate them with a specific substructure, due to the huge dislocation density ($10^{17}$ $m^{-2}$) . It is likely that these atoms are located in such deformed regions that no crystalline substructure can be associated. In this case, these atoms could be seen as being in an almost amorphous configuration.

\newpage

\bibliographystyle{elsarticle-harv} 
\bibliography{ArticleBADAutoencoder2.bib}





% \end{thebibliography}
\end{document}